\documentclass[12pt,preprint]{aastex}
\begin{document}
\title{Cataclysmic Variables from SDSS VII. The Seventh Year (2006)\footnote{Based on 
observations obtained with the Sloan Digital Sky Survey and with the
 Apache Point
Observatory (APO) 3.5m telescope, which are owned and operated by the
Astrophysical Research Consortium (ARC)}}

\author{Paula Szkody\altaffilmark{2},
Scott F. Anderson\altaffilmark{2},
Michael Hayden\altaffilmark{2},
Martin Kronberg\altaffilmark{2},
Rosalie McGurk\altaffilmark{2},
Thomas Riecken\altaffilmark{2},
Gary D. Schmidt\altaffilmark{3},
Andrew A. West\altaffilmark{4},
Boris T. G\"ansicke\altaffilmark{5},
Ada N. Gomez-Moran\altaffilmark{6},
Donald P. Schneider\altaffilmark{7},
Matthias R. Schreiber\altaffilmark{8},
Axel D. Schwope\altaffilmark{6}}

\altaffiltext{2}{Department of Astronomy, University of Washington, Box 351580,
Seattle, WA 98195}
\altaffiltext{3}{The University of Arizona, Steward Observatory, Tucson, AZ 85721}
\altaffiltext{4}{MIT Kavli Institute, 77 Massachusetts Ave, Cambridge MA 02138}
\altaffiltext{5}{Department of Physics, University of Warwick, Coventry CV4 7AL, UK}
\altaffiltext{6}{Astrophysikalisches Institut Potsdam, An der Sternwarte 16, 14482 Potsdam, Germany}
\altaffiltext{7}{Department of Astronomy and Astrophysics, 525 Davey Laboratory, Pennsylvania State University, University Park, PA 16802}
\altaffiltext{8}{Universidad de Valparaiso, Departametno de Fisica y Astronomia, Chile}
\begin{abstract}

Coordinates, magnitudes and spectra are presented for 39 cataclysmic
variables found in Sloan Digital Sky Survey spectra that were primarily obtained in
2006. Of these, 12 were CVs identified prior to the
SDSS spectra (GY Cnc, GO Com, ST LMi, NY Ser, MR Ser, QW Ser, EU UMa, IY UMa, 
HS1340+1524, 
RXJ1610.1+0352, Boo 1, Leo 5).
Follow-up spectroscopic 
 observations of seven systems (including one from year 2005 and another from year 2004) were obtained, 
resulting in estimates of the orbital periods for 3 objects. 
The new CVs include two candidates for high inclination, eclipsing systems,
 4 new Polars and 
three systems whose spectra clearly reveal atmospheric absorption lines
 from the underlying white dwarf. 
\end{abstract}


\section{Introduction}

The Sixth data release from the Sloan Digital Sky Survey 
(SDSS; York et al. 2000) presented the complete photometry of the Galactic
cap as well as further spectroscopy with improved calibrations (Adelman-McCarthy et al. 2008).
Previous releases are detailed by
Stoughton et al. (2002),
Abazajian et al. (2003, 2004,
2005), and Adelman-McCarthy et al. (2006, 2007)\footnote{data are available  
from http://www.sdss.org}. This paper continues the series of identification
of cataclysmic variables (CVs) from the available spectra, with each
paper comprising the objects found in spectra obtained in a given calendar
year (Szkody et al. 2002, 2003, 2004, 2005, 2006, 2007; Papers I-VI). The results for the CVs
found in plates obtained in 2006 are presented here. These objects include
dwarf novae, novalike systems and systems containing highly magnetic
white dwarfs (a comprehensive review of all the various kinds of CVs is contained 
in Warner (1995)). The
number of CVs found in SDSS now constitute a significant sample of uniform
(in resolution and wavelength coverage)
spectra for over 200 objects, and population studies and implications of
the results for different types of CVs are emerging (Schmidt et al. 2005; 
G\"ansicke et al. 2008). While the SDSS is not a targeted CV survey and
not all objects in the photometric sky coverage have spectra obtained to
find CVs, G\"ansicke et al. (200) compare the SDSS sample with the past Palomar
Green and Hamburg Quasar Surveys and consider selection effects. They 
conclude that the primary advantages
of SDSS lie in its great depth and the large amount of spectroscopic
followup of candidates. The increased depth results in a significant
difference in the period distribution found from the SDSS sample of CVs
compared to these previous (brighter) surveys in that the majority of
the SDSS CVs are found at periods below 2 hrs and there is an overabundance
of systems at periods between 80-86 min. This distribution and period spike
follow the predictions of CV evolution models more closely than past surveys.

The above results stem from concentrated efforts by many people in the community
to obtain follow-up photometry and spectroscopy in order to determine the
orbital periods and characteristics of the CVs in the SDSS database (G\"ansicke
et al. (2008) summarize available results for 116, almost half of the total number). 
Our
brief descriptions of the spectra and our few followup observations are
intended to aid these followup studies.

\section{Observations and Reductions}

Detailed information about the SDSS survey (Pier et al. 2003, Gunn et al.
1998, 2006; Lupton, Gunn, \& Szalay 1999;
 Hogg et al. 2001;  Lupton et al. 2001; Ivezic et al. 2004; 
Tucker et al. 2006; Fukugita et al. 1996; Smith et al. 2002; Tucker et al.
2006; Padmanabhan et al. 2008)
and how the CVs are found (Szkody
et al. 2002) from the selection algorithms (Stoughton et al. 2002, Richards
et al. 2002) already
exist in the literature.  
It is important to keep in mind that objects in the imaging data 
are chosen for spectra from
colors that match criteria selected by various
working groups.
CVs are primarily
found that match colors of quasar, serendipity, and white dwarf groups, as the
CVs can be blue if they contain a thick disk, red if they contain
a polar and both red and blue if the disk is thin and the individual stars are
viewed
(typical colors of the CVs found 
in SDSS are plotted in color-color diagrams shown in Papers I,II). While
Table 1 shows that the CVs that do have spectra encompass a wide
range of color, this does not guarantee that all
the CVs in the imaging area covered have spectra obtained.

The search of all spectral plates that are obtained is accomplished via
a software program that selects all objects with Balmer emission/absorption
lines and the selected spectra are visually examined.
All the spectra on a few plates were visually examined to evaluate the
effectiveness of the selection algorithm. While a few are missed if they
are very faint or they are misidentified, we estimate the software finds
about 90\% of the existing CVs.
Table 1 lists the CVs found in SDSS spectra from 2006  
Jan 1 through Dec 31, with  
the plate, fiber, and modified Julian date
(MJD) of each spectrum. There are also a few objects that were missed in
previous years and later recovered. 
 The coordinates are given as equinox J2000.0, with the
IAU convention of truncation rather than rounding at the last decimal, and
the coordinates
have an
 astrometric accuracy of 0.10 arcsec. Photometric magnitudes and colors
are from the 
point-spread function photometry and there is
no correction for interstellar reddening. For ease of reference, we will 
hereafter refer to the objects as 
SDSSJhhmm (hours and min of RA).

For a few objects, we were able to accomplish followup spectroscopy with
the APO 3.5m telescope, using the 
Dual Imaging Spectrograph
(DIS) 
 with the high resolution gratings (resolution about
2\AA) with a 1.5 arcsec slit (Table 2). Two of these followup objects are
from CVs found in previous papers (SDSSJ0812 from Paper V and SDSSJ1006 from
Paper VI). The spectra were obtained over several hours and 
were used to construct
radial velocity curves. 
Calibration for flux and wavelength, as well as measurements of the lines were 
accomplished with standard
IRAF \footnote{{IRAF (Image
 Reduction and Analysis
Facility) is distributed by the National Optical Astronomy Observatories, which
are operated by AURA,
Inc., under cooperative agreement with the National Science Foundation.}}
routines. The SDSS spectra were measured with the centroid-finding 
``e'' routine in the
IRAF $\it{splot}$ package
to obtain the 
 equivalent widths and fluxes for the Balmer and helium emission lines (Table 3). 
For the radial velocity curves, a least squares fit of a sine curve to the velocities
was used to find
 $\gamma$ (systemic velocity), 
K (semi-amplitude), P (orbital period), and $T_{0}$ (the epoch of red
to blue crossing of the systemic velocity);
the results are given in Table 4.
Note that due to the short time baseline of the data, the periods are only 
estimates (with about 10\% accuracy) and will need several nights of further
data for better determinations.
Our measurements,however, provide a starting point as to whether systems have
short or long periods.

\section{Results}

The SDSS spectra for the 39 systems  
are shown in Figure 1 and 
the equivalent widths and fluxes of the prominent hydrogen Balmer
and helium emission lines are listed in Table 3. A summary of the
various categories of objects is given below.  

\subsection{Previously Known Systems}

Of the 39 entries in Table 1, 12 are CVs that were found prior to SDSS
spectra. These include the novalike Leo5 (1H1025+220;SDSSJ1029) and seven
 dwarf novae: Boo1 (SDSSJ1504), GY Cnc (SDSSJ0909), GO Com (SDSSJ1256), 
 NY Ser (SDSSJ1513), QW Ser (SDSSJ1526), IY UMa (SDSSJ1043) and
 HS1340+1524 (SDSSJ1343). Leo5 was previously identified as a CV candidate
during follow-up of HEAO-1 sources, and spectroscopically confirmed by
Munari \& Zwitter(1998).
Boo1 was discovered as a faint emission line
star by Filipenko et al. (1985), who tentatively classified the
object as a dwarf nova, even though no outburst was observed, and
no follow-up observations have been obtained so far.
GY\,Cnc was identified as an eclipsing dwarf nova with
$P_\mathrm{orb}=252.6$\,min; an updated ephemeris is given by
Feline et al. (2005).
GO\,Com has been long known as a dwarf nova
(Brun \& Petit 1957), and had $P_\mathrm{orb}=95$\,min determined by
Howell et al. (1995). 
NY\,Ser was identified as a CV in the Palomar-Green Survey
(Green et al. 1986), and a short outburst cycle was noted by
Iida et al. (1995). Nogami et al. (1998) measured a superhump
period of 153\,min, making NY\,Ser the first SU\,UMa type dwarf nova
in the period gap between 2-3 hrs where few CVs are found (Warner 1995). 
Patterson et al. (2003) determined the orbital
period as 140.4\,min. 
QW\,Ser was identified as a dwarf nova by
Takamizawa (1998), and $P_\mathrm{orb}=107.3$\,min was determined
by Patterson et al. (2003). 
IY\,UMa is another eclipsing dwarf nova with an $P_\mathrm{orb}=106.4$\,min
(Uemura et al. 2000), for an updated ephemeris see
Steeghs et al. (2003).
HS1340+1524 (SDSSJ1343) is a dwarf nova with infrequent short
outbursts, and $P_\mathrm{orb}=92.7$\,min
(Aungwerojwit et al. 2006). 

Finally, there are four previously known polars among the SDSS CVs
presented here. Two of them were observed in a low state: ST\,LMi (SDSSJ1105), one
of the few polars identified in the optical (Shore et al. 1982) 
with $P_\mathrm{orb}=113.9$\,min (Schmidt et al. 1983, Cropper 1986); and
EU\,UMa (SDSSJ1149), discovered with ROSAT (Mittaz et al. 1992) with
$P_\mathrm{orb}=90$\,min (Howell et al. 1995). The two other polars
were observed by SDSS during high states, MR\,Ser (SDSSJ1552),
identified in the PG survey (Liebert et al. 1982) with
$P_\mathrm{orb}=113.5$\,min (Schwope et al. 1991), and
RXS\,J161008.0+035222 (SDSSJ1610), identified as a ROSAT polar by
Schwope et al. (2000, 2002), with recent polarimetry
published by Rodrigues et al. (2006) which refined the orbital period to
109.5 min. 

 Table 1 also includes four Polars that we found since Paper
VI which have detailed information recently published 
(Schmidt et al. 2007, 2008); we include them in the Table for completeness:
SDSSJ0921, SDSSJ1031, SDSSJ1059, and SDSSJ1333. Of these,
SDSSJ1031 and SDSSJ1059 belong to the group of extremely low mass transfer
rate polars, while the rest are normal polars with high and low states
of accretion. Note that the magnitudes listed for SDSSJ0921
and SDSSJ1333 in Schmidt et al. 2008 are in juxtaposed order in their
Table (the magnitudes are actually in order of $\it{g,i,r,u,z}$ instead
of $\it{u,g,r,i,z}$ as labeled.

\subsection{High Inclination Systems}

Previous work on SDSS systems has shown that those with deep central
absorption in the Balmer lines typically have high inclination
and show photometric eclipses. Two systems, SDSSJ1057 and SDSSJ1524 (Figure 1),
show this central absorption,
and are promising candidates for
having deep eclipses of the white dwarf by the secondary star.

\subsection{Dwarf Novae}

While CVs can be generally identified by their emission line spectrum,
the identification of a dwarf nova requires that an outburst is apparent.
This can be apparent from a difference in the SDSS photometry versus the
spectra (which are obtained at different times) or as large differences
in magnitude in past USNO or DSS catalogs or in other non-SDSS observations.
The known dwarf nova QW Ser (SDSSJ1526) was caught at outburst in the
SDSS spectra (Figure 1) while the photometry (Table 1) is consistent with
its normal quiescent magnitude near 18. 

{\bf{\it SDSSJ1005:}} A report of
an outburst of this object by Brady \& Pietz (2009) 
recently appeared in the vsnet  
\footnote{http://vsnet.kusastro.kyoto-u.ac.jp/vsnet/}, thus providing
a classification for this system. Subsequent searches of ASAS-3 data
as reported by Kato (2009) showed previous
outbursts near 12.5mag in 2003 and 2006. 

  Our followup APO time-resolved spectra during quiescence in 2007
produced consistent results from the H$\alpha$ and H$\beta$ emission
lines. The period obtained from both lines is near 113 min and the K
amplitude is low (Figure 3 and Table 4). While further data over several
nights will be needed to pin this down precisely, it is apparent that this
is likely a low inclination, short period system that is near the lower
edge of the period gap. The preliminary superhump period reported by
Brady \& Pietz
is identical to our spectroscopic period within the accuracy
reported.

{\bf{\it SDSSJ1619:}} The SDSS photometry (Table 1) and spectrum (Figure 1)
show a typical CV at quiescence, with an optical magnitude near 18.5 and
Balmer emission lines with a flat decrement. However, our followup APO
spectra (Table 2 and Figure 2) show a much brighter source (near magnitude
15.5) with strong \ion{He}{2}4686 emission as well as weaker Balmer emission
flanked by broad absorption. The APO spectra are typical of dwarf novae
at outburst, where the increased accretion at outburst results in the
high excitation He line and an optically thick accretion disk which produces
the broad absorption. Thus, we can narrow the classification of this object
to that of dwarf nova. Our time-resolved APO data covered close to two
hours of observation, but our measurements of the H$\alpha$, H$\beta$ and
\ion{He}{2} emission components did not reveal any periodic radial velocity
variation outside of random variability that was $<$ 20 km/s. Thus, either
this system has a low inclination, a long period, or the emission lines
at outburst are too distorted by the underlying absorption to extract the
underlying orbital motion. Further observations during quiescence are needed
to determine its orbital period.

{\bf{\it SDSSJ1627:}} A superoutburst has recently been detected by
Shears et al. (2008), who determined a superhump period of
$P_\mathrm{sh}=156.8$\,min. Since the superhump period is usually only a
few percent different from the orbital period (Warner 1995), this system
appears to be one of the few in the $2-3$ orbital
period gap.

\subsection{Nova-likes with \ion{He}{2}}

The \ion{He}{2}4686 line is a strong indicator of a polar or of high accretion. All of
the polars mentioned in section 3.1 show this line (except for the two
with extremely low accretion rates). In addition to these known polars,
Figure 1 reveals three other systems with unusually strong \ion{He}{2}4686.

{\bf{\it SDSSJ1549:}} This object has a very peculiar spectrum, showing a
strong continuum, weak Balmer emission but very strong \ion{He}{2}. The SDSS
spectrum is very similar to that of UMa 6 (SDSSJ0932) shown in Paper V.
UMa 6 has a very long orbital period for a CV 
(10 hrs\footnote{http://cbastro.org/results/highlights/uma6}) and a deep 
optical eclipse (Hilton et al. 2008). Our 2.5 hrs of APO time-resolved
spectroscopy (Table 2) showed 40 km/s variability in both H$\alpha$
and \ion{He}{2} but no simple sinusoidal motion consistent with an
orbital radial velocity. Thus, this system will require much longer
monitoring to ascertain its nature.

{\bf{\it SDSSJ0938:}} The spectrum of SDSSJ0938 looks typical for a polar
(Figure 1) in a high state of accretion. It is virtually identical to the known polar SDSSJ1610 also
in Figure 1. Spectropolarimetry will be able to provide definitive information
on this issue.
While our APO observations (Table 2) were not long enough to obtain an
orbital period, a smooth, large amplitude (70 km/s) variation throughout
the 65 min is consistent with a polar with a period that is under 2 hrs.

{\bf{\it SDSSJ0935:}} While this object has stronger \ion{He}{2} than
H$\beta$ emission (Figure 1 and Table 3), the spectral appearance is
different than for the above two objects. The continuum is very strong
and the emission lines are broad and weak. This spectrum appears more like
an old nova than a system containing a magnetic white dwarf (Warner 1995).

\subsection{Systems Showing the Underlying Stars}

The ability of SDSS to obtain spectra of CVs that are fainter than previous
surveys has resulted in discovering many systems that have low accretion
rates, hence accretion disks which do not overwhelm the light of the
underlying stars. In these cases, the white dwarfs are revealed through
their broad absorption lines flanking the Balmer emission and, if the
secondary star is a late main sequence object, it is
evident by TiO features in the red. From Figure 1, it is apparent that
SDSSJ1005, SDSSJ1057 and SDSSJ1605 show the white dwarf, while SDSSJ0230,
SDSSJ1059, SDSSJ1105, SDSSJ1544 show an M star (SDSSJ1105 and SDSSJ1059
are known polars with no accretion disk) and SDSSJ0805 appears to show
a K star (albeit of somewhat later type than the K stars in SDSSJ0615
and SDSSJ0805 found in Paper VI).

\subsection{Other Disk Systems}

The spectra of systems with accretion disks can show a large range in variety
(Warner 1995). Figure 1 shows five systems with strong, blue
continua: SDSSJ0758, SDSSJ0901, SDSSJ0935 (already mentioned in section 3.4),
SDSSJ1054, and SDSSJ1513. Most of these are likely to be novalikes with
large accretion rates. SDSSJ1054 may be questionable as it could be just
a white dwarf and a faint active but non-interacting M star. Two spectra
taken 26 days apart exist in the SDSS archive for this object and they show 
minor differences
in the structure of the Balmer emission and absorption lines which could
be due to a close binary so we have left this in the list. Followup
spectroscopy will determine the correct classification. 
The systems with weaker continua and stronger emission
lines are likely candidates for short orbital period systems with lower
mass transfer. For our followup APO spectra, we generally concentrated on
these latter systems due to the way observing time is scheduled in half-nights.

{\bf{\it SDSSJ1557:}} This object has strong, broad Balmer emission lines 
that are typical for dwarf novae systems. Our 2.5 hrs of time-resolved
spectra revealed a sinusoidal modulation with a period near 2 hrs, which is
the lower
end of the period gap (Table 4 and Figure 4). The amplitude is typical
for dwarf novae. The object will need to be followed photometically
to detect an outburst and confirm this as a dwarf nova.

{\bf{\it SDSSJ0812:}} Followup 3.5 hrs of time-resolved spectra of this CV that was first reported in
Paper V shows a high amplitude radial velocity curve with a
period near 3.7 hrs, close to the length of the dataset (Table 4 and Figure 5). This object  
thus appears to be above the period gap and have a higher accretion rate than
the majority
of SDSS CVs that have periods less than 2 hrs.

{\bf{\it SDSSJ1006:}} This system from paper VI was targeted for followup
spectra as it shows strong emission lines plus TiO bands from its secondary
star. However, 100 min of spectra do not reveal a clear sinusoidal
variation. There is a jump in velocities in both H$\alpha$ and H$\beta$ from
red to blue (with no change in comparison lamps taken near these times)
and there is a decline in flux in the spectra at this time. These
properties could be an indication of an eclipse, so additional data on this
object could produce interesting results.

\subsection{ROSAT Correlations}

Ten of the objects in Table 1 have been detected with the
ROSAT All Sky Survey (RASS; Voges et al. 1999, 2000). The exposure times
and count rates are listed in Table 5. Among the 10 detections are
the known polars EU UMa, MR Ser and RXJ1610+03 and the dwarf novae GY Cnc, GO Com,
QW Ser and HS1340+15. Several of the sources are only in the faint source
catalog with marginal detections (no error is listed for the faint detection
of SDSSJ1005). The lack of detection of the other polars such as ST LMi
and the LARPS detailed in Schmidt et al. (2007, 2008)
 are indications of states of very low mass
transfer for those systems.  On the other hand, the detection of SDSSJ0938
lends further support for this object being a possible polar. As in UMa 6, 
the strong
HeII present in SDSSJ1549 
is not correlated with X-ray emission.
 
\section{Conclusions}

The addition of these 39 objects to the previous list brings
 the total number of CVs in
the SDSS database to 252, of which 204 are new discoveries.
There are now more than 100 CVs with known or estimated orbital periods (see
G\"ansicke et al. 2008 for a recent summary). The distribution of periods
of objects from SDSS is significantly different than previous surveys with
brighter limits. The SDSS objects exist predominantly at short periods
and show a period spike at 81 min, as predicted by binary evolution
theories. Thus, this database can serve as a testbed for evolution and
further period determinations will refine these numbers.

The following objects should be of high interest for future studies.
Followup photometry, spectroscopy and especially polarimetry of SDSJS0938
will confirm if this system contains a magnetic white dwarf.
Photometry of SDSSJ1057 and SDSSJ1524 is likely to reveal eclipses which
can determine inclinations and periods. SDSSJ1006 from Paper VI may
also have eclipses.
 High time-resolution photometry of SDSSJ1005,
SDSSJ1057 and SDSSJ1605 should be done to search for pulsations of the white dwarf.
Long term photometry of SDSSJ1549 is needed to determine if the large
differences in magnitude that are apparent are due to a long orbital period
with eclipses (like UMa 6) or different states of low and high accretion.
Spectroscopy (especially in the IR) for the two systems showing indications
of the secondary star (SDSSJ0230 and SDSSJ1544) can produce better information
on the secondary and the likely longer orbital periods in these two systems.

\acknowledgments

    Funding for the SDSS and SDSS-II has been provided by the Alfred P. Sloan 
Foundation, the Participating Institutions, the National Science Foundation, the
 U.S. Department of Energy, the National Aeronautics and Space Administration, 
the Japanese Monbukagakusho, the Max Planck Society, and the Higher Education 
Funding Council for England. The SDSS Web Site is http://www.sdss.org/.

    The SDSS is managed by the Astrophysical Research Consortium for the 
Participating Institutions. The Participating Institutions are the American 
Museum of Natural History, Astrophysical Institute Potsdam, University of Basel,
 University of Cambridge, Case Western Reserve University, University of 
Chicago, Drexel University, Fermilab, the Institute for Advanced Study, the 
Japan Participation Group, Johns Hopkins University, the Joint Institute for 
Nuclear Astrophysics, the Kavli Institute for Particle Astrophysics and 
Cosmology, the Korean Scientist Group, the Chinese Academy of Sciences (LAMOST),
 Los Alamos National Laboratory, the Max-Planck-Institute for Astronomy (MPIA),
 the Max-Planck-Institute for Astrophysics (MPA), New Mexico State University, 
Ohio State University, University of Pittsburgh, University of Portsmouth, 
Princeton University, the United States Naval Observatory, and the University 
of Washington.

P.S. acknowledges support from NSF grant AST 0607840.
Studies of magnetic stars and stellar systems at Steward Observatory is 
supported by the NSF through AST 03-06080. M.R.S. acknowledges support
from FONDECYT (grant 1061199).

\clearpage

\begin{deluxetable}{lccrrrrll}
\tabletypesize{\scriptsize}
\tablewidth{0pt}
\tablecaption{CVs in SDSS}
\tablehead{
\colhead{SDSS J} &  \colhead{MJD-P-F\tablenotemark{a}} & 
\colhead{$g$} & \colhead{$u-g$} & \colhead{$g-r$} & 
\colhead{$r-i$} &
\colhead{$i-z$} & \colhead{P(hr)} & \colhead{Comments\tablenotemark{b}} }
\startdata
023003.79+260440.3 & 53764-2399-405 & 19.91 & 0.29 & 0.72 & 0.60 & 0.43 & ... & \\
075808.81+104345.5 & 53794-2418-278 & 16.96 & 0.14 & -0.13 & -0.10 & -0.07 & ... & \\
082253.12+231300.6 & 53317-1926-544 & 21.84 & 0.76 & 1.93 & 0.65 & 0.31 & ... &
 \\
085521.18+111815.0 & 54085-2575-318 & 18.81 & 0.13 & 0.06 & 0.15 & 0.31 & ... & \\
090113.51+144704.6 & 53826-2434-400 & 16.14 & 0.21 & -0.02 & -0.07 & -0.07 & ... & \\
090950.53+184947.3 & 53687-2285-030 & 16.05 & -0.13 & 0.34 & 0.39 & 0.31 & 4.21 & GY Cnc \\
091001.63+164820.0 & 53828-2435-075 & 18.87 & -0.34 & 0.30 & 0.01 & 0.22 & ... & \\
092122.84+203857.1 & 53708-2289-316 & 19.85 & 0.79 & 0.68 & -0.01 & -0.35 & $>$1.5 & Polar \\
093537.46+161950.8 & 54085-2581-332 & 19.10 & 0.42 & 0.08 & -0.01 & -0.01 & ... & HeII \\
093839.25+534403.8 & 53764-2404-414 & 19.15 & 0.94 & 0.31 & 0.19 & 0.02 & ... & HeII \\
100515.38+191107.9 & 53768-2372-473 & 18.22 & -0.07 & -0.05 & 0.15 & 0.26 & 1.9 & DN \\
102800.08+214813.5 & 53741-2366-072 & 16.06 & 0.37 & -0.07 & -0.10 & -0.09 & ... & 1H1025+220 Leo 5\\
103100.55+202832.2 & 53770-2375-636 & 18.26 & 0.09 & -0.28 & -0.36 & 0.15 & 1.37 & Polar \\
104356.72+580731.9 & 52427-0949-0358 & 17.52 & 0.18 & -0.10 & 0.02 & 0.45 & 1.77 & IY UMa \\
105443.06+285032.7 & 53800-2359-497 & 19.23 & -0.32 & -0.49 & -0.15 & 0.14 & ... &  \\
105754.25+275947.5 & 53800-2359-102 & 19.90 & -0.30 & 0.27 & -0.18 & 0.16 & ... & \\
105905.07+272755.5 & 53800-2359-051 & 22.09 & 1.27 & 1.84 & 0.35 & 1.05 & $>$3 & Polar \\
110539.76+250628.6 & 53789-2212-201 & 17.63 & 0.39 & 0.04 & 0.57 & 0.76 & 1.90 & ST LMi Polar\\
114955.69+284507.3 & 53799-2222-010 & 17.63 & -0.05 & -0.06 & -0.11 & 0.27 & 1.50 & EU UMa Polar \\
124417.89+300401.0 & 53828-2237-560 & 18.61 & -0.03 & 0.10 & 0.10 & 0.27 & ... & \\
125637.10+263643.2 & 53823-2240-092 & 17.98 & 0.06 & 0.07 & 0.02 & 0.18 &  1.58 & GO Com \\
133309.19+143706.9 & 53847-1775-428 & 18.50 & 0.57 & 0.36 & 0.18 & 0.03 & 2.2 & Polar \\
134323.16+150916.8 & 53858-1776-576 & 17.34 & -0.36 & 0.18 & 0.04 & 0.06 & 1.54 & HS1340+1524 \\
150441.76+084752.6 & 53883-1717-260 & 19.14 & -0.54 & -0.02 & 0.06 & 0.35 &  ... & Boo 1 \\ 
151302.29+231508.4 & 53820-2155-163 & 16.09 & 0.16 & -0.13 & -0.04 & -0.03 & 2.35 & NY Ser \\
152212.20+080340.9 & 53857-1721-209 & 18.42 & -0.14 & -0.02 & 0.02 & 0.15 & ...  & \\
152419.33+220920.0 & 53878-2161-189 & 19.04 & -0.03 & 0.16 & 0.09 & 0.31 & ... & \\
152613.96+081802.3 & 53857-1721-021 & 17.79 & 0.00 & -0.02 & 0.13 & 0.29 & 1.79 & QW Ser \\
153015.04+094946.3 & 53852-1722-141 & 18.90 & -0.49 & 0.41 & 0.02 & -0.06 & ... & \\
154453.60+255348.8 & 53846-1849-074 & 16.60 & -0.13 & 0.46 & 0.15 & 0.34 & ... & \\
154953.41+173939.0 & 53875-2170-276 & 19.44 & 0.31 & 0.39 & 0.18 & 0.01 & ... & \\
155247.18+185629.1 & 53875-2170-441 & 17.21 & 0.21 & -0.11 & 0.29 & 0.69 & 1.89 & MR Ser Polar \\
155720.75+180720.2 & 53875-2170-588 & 18.70 & -0.58 & 0.22 & 0.15 & 0.10 & 2.1 & ...\\ 
160419.02+161548.5 & 53875-2200-292 & 19.09 & -0.37 & 0.26 & 0.07 & 0.05 & ... & \\
160501.35+203056.9 & 53793-2205-247 & 19.89 & -0.10 & 0.01 & -0.07 & -0.17 & ...  &  \\
160932.67+055044.6 & 53886-1823-411 & 18.77 & 0.12 & -0.07 & -0.10 & -0.04 & ... & \\
161007.50+035232.7 & 53886-1823-092 & 17.36 & -0.25 & 0.15 & 0.41 & 0.47 & 3.18 & Polar \\
161909.10+135145.5 & 53881-2530-327 & 18.49 & 0.43 & 0.68 & 0.39 & 0.27 & ... & DN \\
162718.39+120435.0 & 53881-2530-068 & 19.22 & -0.23 & 0.17 & 0.23 & 0.37 & 
2.61\tablenotemark{c} & DN \\
\enddata
\tablenotetext{a}{MJD-Plate-Fiber for spectra; MJD = JD - 2,400,000.5}
\tablenotetext{b}{DN is a dwarf nova}
\tablenotetext{c}{superhump period}
\end{deluxetable}

\clearpage
\begin{deluxetable}{lcccc}
\tablewidth{0pt}
\tablecaption{APO Follow-up Spectroscopy}
\tablehead{
\colhead{SDSSJ} & \colhead{UT Date} & 
\colhead{Time (UT)} & \colhead{Exp (s)} & \colhead{Spectra} }
\startdata
1549 & 2006 Jun 17 & 04:21-06:56 & 600 & 14 \\
0812\tablenotemark{a} & 2006 Oct 22 & 08:55-12:26 & 600 & 17 \\
1005 & 2007 Apr 20 & 03:18-05:54 & 600 & 14 \\
1619 & 2007 May 10 & 04:43-06:43 & 600 & 11 \\
1006\tablenotemark{b} & 2007 May 10 & 02:48-04:31 & 600 & 09 \\
1557 & 2007 Jul 19 & 03:43-06:24 & 600 & 14 \\
0938 & 2008 Jan 16 & 05:56-07:10 & 600 & 07 \\
\enddata
\tablenotetext{a}{object discovered in Paper V}
\tablenotetext{b}{object discovered in Paper VI}
\end{deluxetable}

\clearpage
\begin{deluxetable}{lrrrrrrrr}
\tabletypesize{\scriptsize}
\tablewidth{0pt}
\tablecaption{SDSS Emission Line Fluxes and Equivalent Widths\tablenotemark{a}}
\tablehead{
\colhead{SDSSJ} &  
\multicolumn{2}{c}{H$\beta$} &
\multicolumn{2}{c}{H$\alpha$} &
\multicolumn{2}{c}{He4471} & \multicolumn{2}{c}{HeII4686}\\
\colhead{} &  \colhead{F} &
\colhead{EW} & \colhead{F} & \colhead{EW} & \colhead{F} & \colhead{EW} &
\colhead{F} & \colhead{EW} } 
\startdata
0238 & 2.6 & 65 & 4.4 & 74 & 0.7 & 16 & ... & ... \\
0758 & 2.3 & 4 & 3.2 & 10 & ... & ... & ... & ... \\
0822 & ... & ... & 0.8 & 25 & ... & ... & ... & ... \\
0855 & 4.0 & 31 & 5.5 & 72 & ... & ... & ... & ... \\
0901 & 2.1 & 2 & 3.4 & 4 & ... & ... & ... & ... \\
0909 & 74.6 & 57 & 108.0 & 85 & 13.6 & 11 & 5.5 & 4 \\
0910 & 8.7 & 89 & 10.2 & 154 & 1.7 & 17 & ... & ... \\
0921 & 1.6 & 15 & 1.4 & 17 & 0.6 & 6 & 0.4 & 4 \\
0935 & 0.6 & 4 & 1.2 & 16 & ... & ... & 1.1 & 7 \\
0938 & 2.1 & 18 & 2.8 & 29 & 0.7 & 7 & 1.5 & 13 \\
1005 & 7.5 & 45 & 10.4 & 94 & 1.1 & 4 & ... & ... \\
1028 & 5.4 & 5 & 9.4 & 14 & ... & ... & 0.6 & 1 \\
1031 & ... & ... & 0.3 & 6 & ... & ... & ... & ... \\
1043 & 3.7 & 7 & 9.9 & 35 & ... & ... & ... & ... \\
1054 & 0.1 & 2 & 0.2 & 7 & ... & ... & ... & ... \\
1057 & 1.1 & 22 & 2.6 & 95 & ... & ... & ... & ... \\
1059 & ... & ... & 4.0 & 2 & ... & ... & ... & ... \\
1105 & 3.2 & 9 & 0.7 & 3 & ... & ... & ... & ... \\
1149 & 0.1 & 1 & 0.4 & 7 & ... & ... & ... & ... \\
1244 & 12.0 & 85 & 13.1 & 115 & 2.4 & 15 & ... & ... \\
1256 & 37.4 & 101 & 35.1 & 124 & 6.5 & 15 & 4.9 & 13 \\
1333 & 1.6 & 150 & 1.2 & 107 & 0.3 & 19 & 1.0 & 77 \\
1343 & 20.0 & 71 & 21.2 & 100 & 4.2 & 14 & 1.7 & 6 \\
1504 & 12.3 & 114 & 12.5 & 165 & 2.7 & 22 & 0.6 & 5 \\
1513 & 1.0 & 9 & 1.6 & 20 & ... & ... & ... & ... \\
1522 & 7.1 & 108 & 8.7 & 182 & 1.5 & 22 & 0.9 & 14 \\
1524 & 3.6 & 59 & 5.1 & 123 & ... & ... & ... & ... \\
1526 & ... & ... & ... & ... & ... & ... & ... & ... \\
1530 & 5.3 & 56 & 4.9 & 70 & 1.3 & 13 & 0.6 & 7 \\
1544 & 34.8 & 65 & 46.7 & 77 & 11.1 & 21 & 5.1 & 10 \\
1549 & 0.9 & 2 & 1.8 & 6 & ... & ... & 5.5 & 10 \\
1552 & 60.5 & 27 & 78.5 & 33 & 26.2 & 12 & 33.8 & 16 \\
1557 & 13.4 & 64 & 11.3 & 71 & 3.0 & 12 & 1.4 & 6 \\
1604 & 5.2 & 63 & 5.5 & 97 & 1.2 & 14 & ... & ... \\
1605 & 1.1 & 19 & 2.3 & 78 & ... & ... & ... & ... \\
1609 & 0.7 & 12 & 1.1 & 36 & ... & ... & ... & ... \\
1610 & 46.5 & 35 & 36.8 & 27 & 11.3 & 9 & 31.0 & 25 \\
1619 & 7.3 & 27 & 8.8 & 33 & 1.1 & 4 & ... & ... \\
1627 & 7.5 & 83 & 7.1 & 94 & 2.7 & 28 & 1.7 & 19 \\
\enddata
\tablenotetext{a}{Fluxes are in units of 10$^{-15}$ ergs cm$^{-2}$ s$^{-1}$,
equivalent widths are in units of \AA}
\end{deluxetable}

\clearpage
\normalsize
\begin{deluxetable}{lcccccc}
\tablewidth{0pt}
\tablecaption{Radial Velocity Solutions}
\tablehead{
\colhead{SDSSJ} & \colhead{Line} & \colhead{P (min)\tablenotemark{a}} & 
\colhead{$\gamma$} & \colhead{K (km s$^{-1}$)} &
\colhead{T$_{0}$ (JD2,454,000+)} & \colhead{$\sigma$} }
\startdata
0812 & H$\alpha$ & 229 & -30$\pm$1 & 208$\pm$11 & 30.989 & 27 \\
0812 & H$\beta$ & 215 & -38$\pm$1 & 172$\pm$13 & 30.988 & 35 \\
1005 & H$\alpha$ & 112 & 42$\pm$1 & 19$\pm$4 & 210.658 & 10 \\
1005 & H$\beta$ & 114 & 38$\pm$1 & 27$\pm$3 & 210.663 & 7 \\
1557 & H$\alpha$ & 122 & -50$\pm$3 & 89$\pm$13 & 300.720 & 34 \\
1557 & H$\beta$  & 133 & -43$\pm$2 & 117$\pm$13 & 300.727 & 33 \\
\enddata
\tablenotetext{a}{Periods are generally uncertain by 
10\%, as evidenced by the dispersion between values obtained from the 2 
lines.} 
\end{deluxetable}

\clearpage
\begin{deluxetable}{lccll}
\tablewidth{0pt}
\tablecaption{ROSAT Detections}
\tablehead{
\colhead{SDSSJ} & \colhead{ROSAT (c s$^{-1}$)\tablenotemark{a}} & \colhead{Exp (s)}
& \colhead{RXS} & \colhead{Type} }
\startdata
0909 & 0.08$\pm$0.02 & 364 & J090950.6+184956=GY Cnc & DN \\
0938 & 0.03$\pm$0.01 & 409 & J093838.0+534417 & ... \\
1005 & 0.03 & 414 & J100511.9+191105 & DN \\
1149 & 3.33$\pm$0.16 & 127 & J114955.5+284510=EU UMa & Polar \\
1256 & 0.06$\pm$0.01 & 476 & J125637.6+263656=GO Com & DN \\
1343 & 0.07$\pm$0.02 & 354 & J134323.1+150916  & DN \\
1526 & 0.05$\pm$0.02 & 277 & J152613.9+081845=QW Ser & DN \\
1552 & 0.04$\pm$0.01 & 595 & J155246.3+185608=MR Ser & Polar \\
1557 & 0.014$\pm$0.007 & 587 & J155720.3+180715 & ... \\
1610 & 0.36$\pm$0.04 & 494 & J161008.0+035222 & Polar \\
\enddata
\tablenotetext{a}{For a 2 keV bremsstrahlung spectrum, 1 c s$^{-1}$ corresponds to a
0.1-2.4 keV flux of about 7$\times10^{-12}$ ergs cm$^{-2}$ s$^{-1}$}
\end{deluxetable}

\clearpage

\begin{figure} [p]
\figurenum {1a}
\plotone{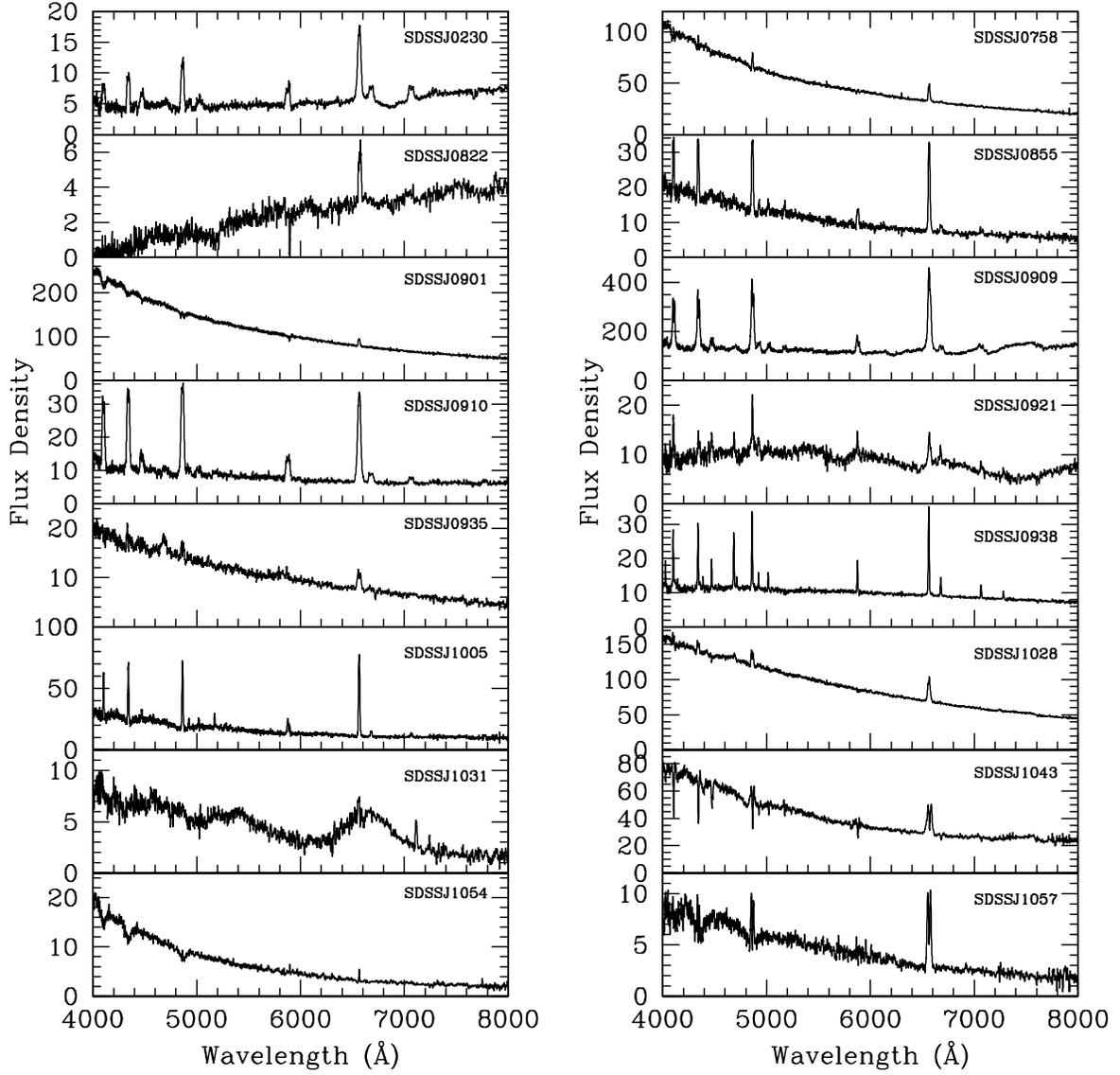}
\caption{SDSS spectra of the 36 CVs.
 Vertical axis is 
units of
flux density F$_{\lambda}\times$10$^{-17}$ ergs cm$^{-2}$ s$^{-1}$ \AA$^{-1}$. The spectral
resolution is about 3\AA.}
\end{figure}

\begin{figure} [p]
\figurenum {1b}
\plotone{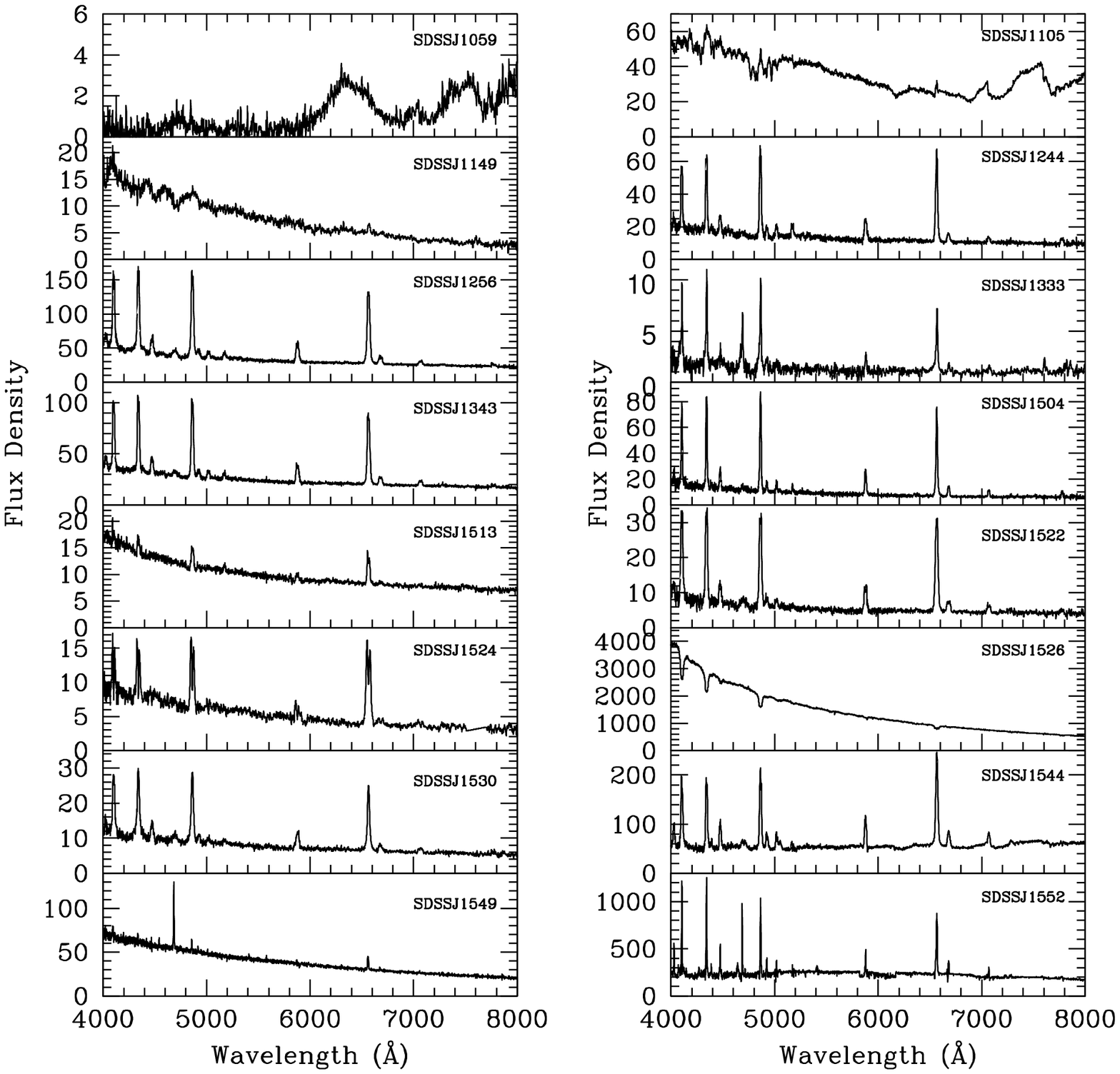}
\caption{Continued.}
\end{figure}

\begin{figure} [p]
\figurenum {1c}
\plotone{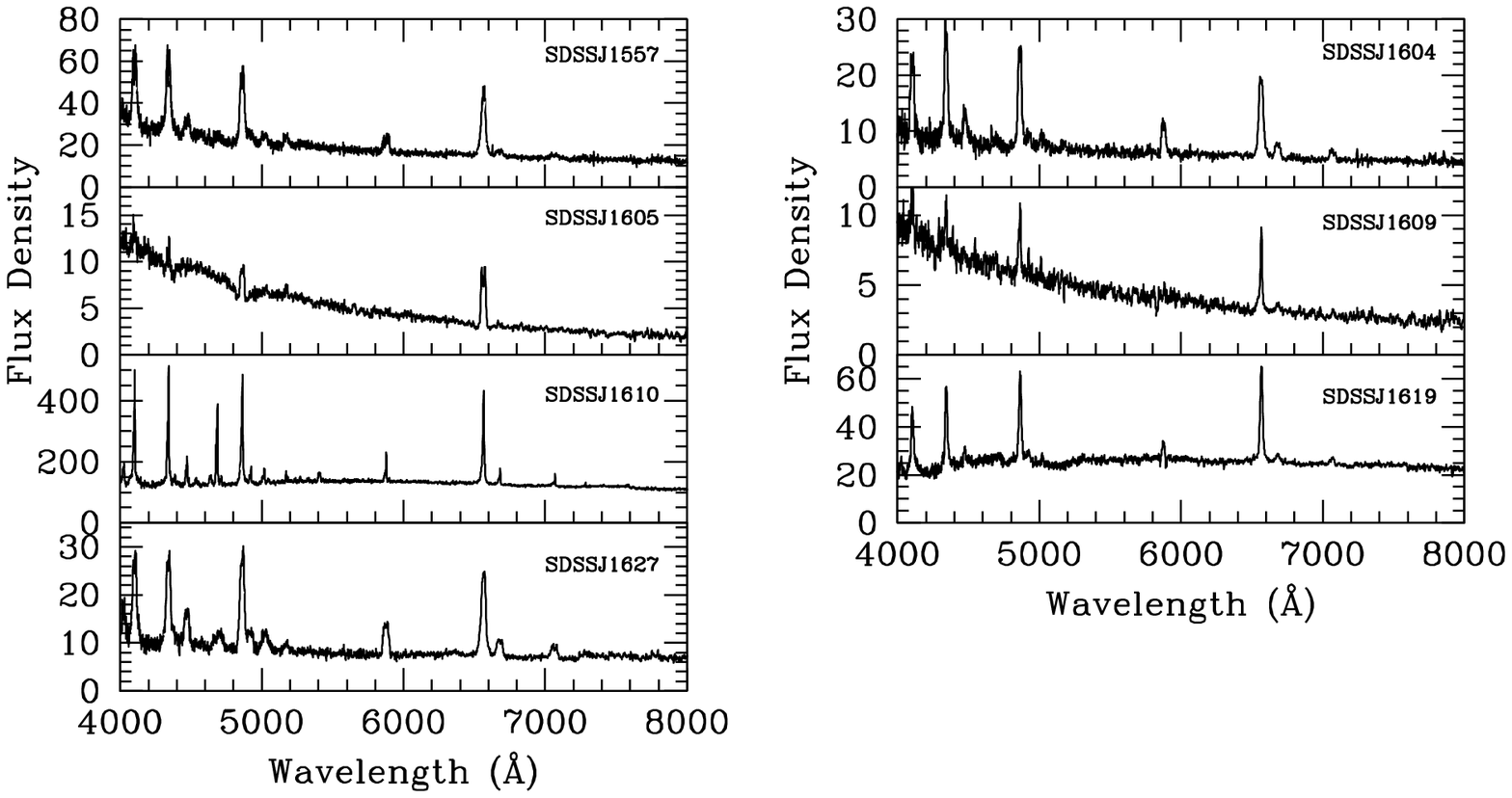}
\caption{Continued.}
\end{figure}

\begin{figure} [p]
\figurenum {2}
\plotone{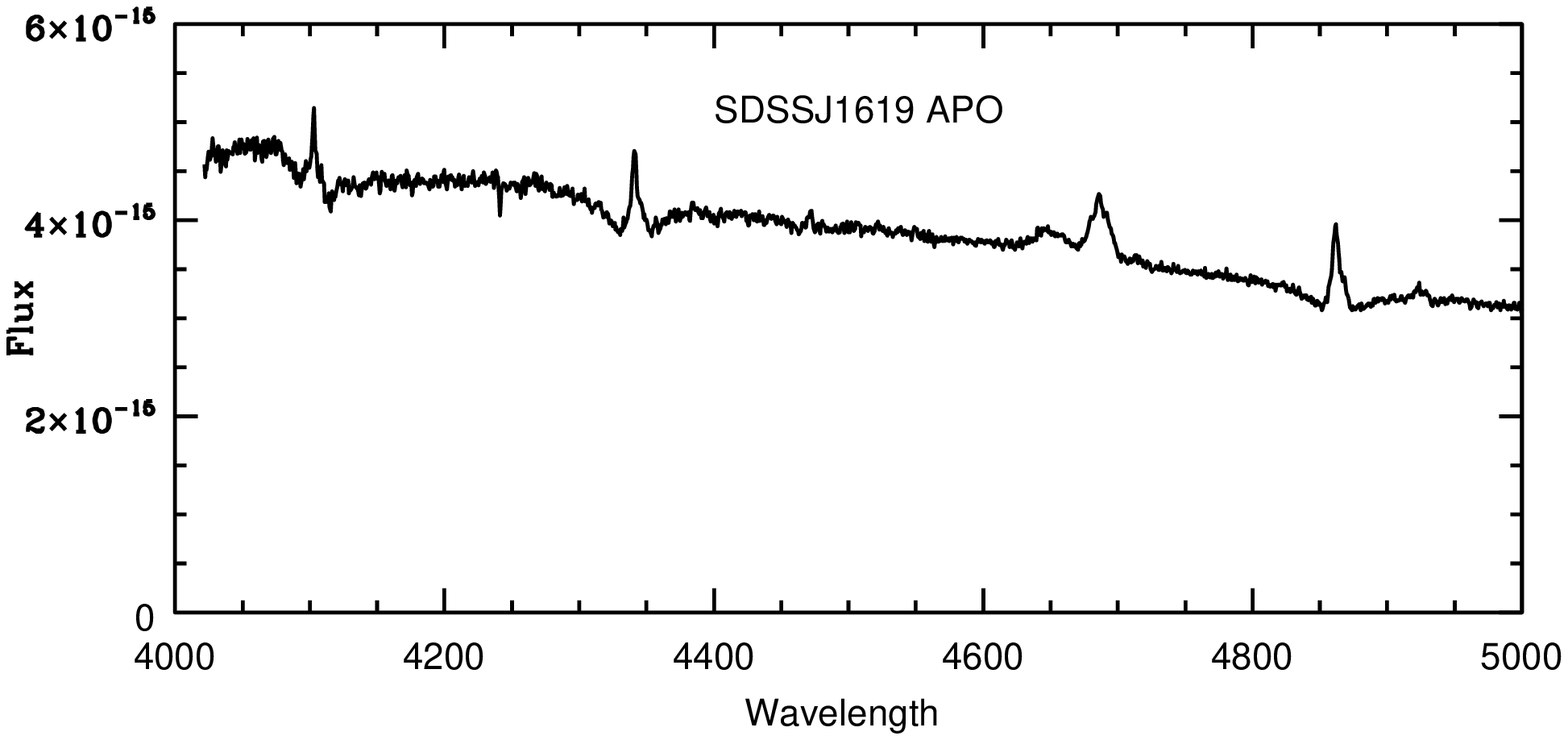}
\caption{Combined APO spectra of SDSSJ1619 obtained on 2007 May 10 during an outburst. Note increased
HeII emission, Balmer absorption, and higher flux as compared to spectrum in Figure 1. The spectral resolution is about 2\AA.}
\end{figure}

\begin{figure} [p]
\figurenum {3}
\plotone{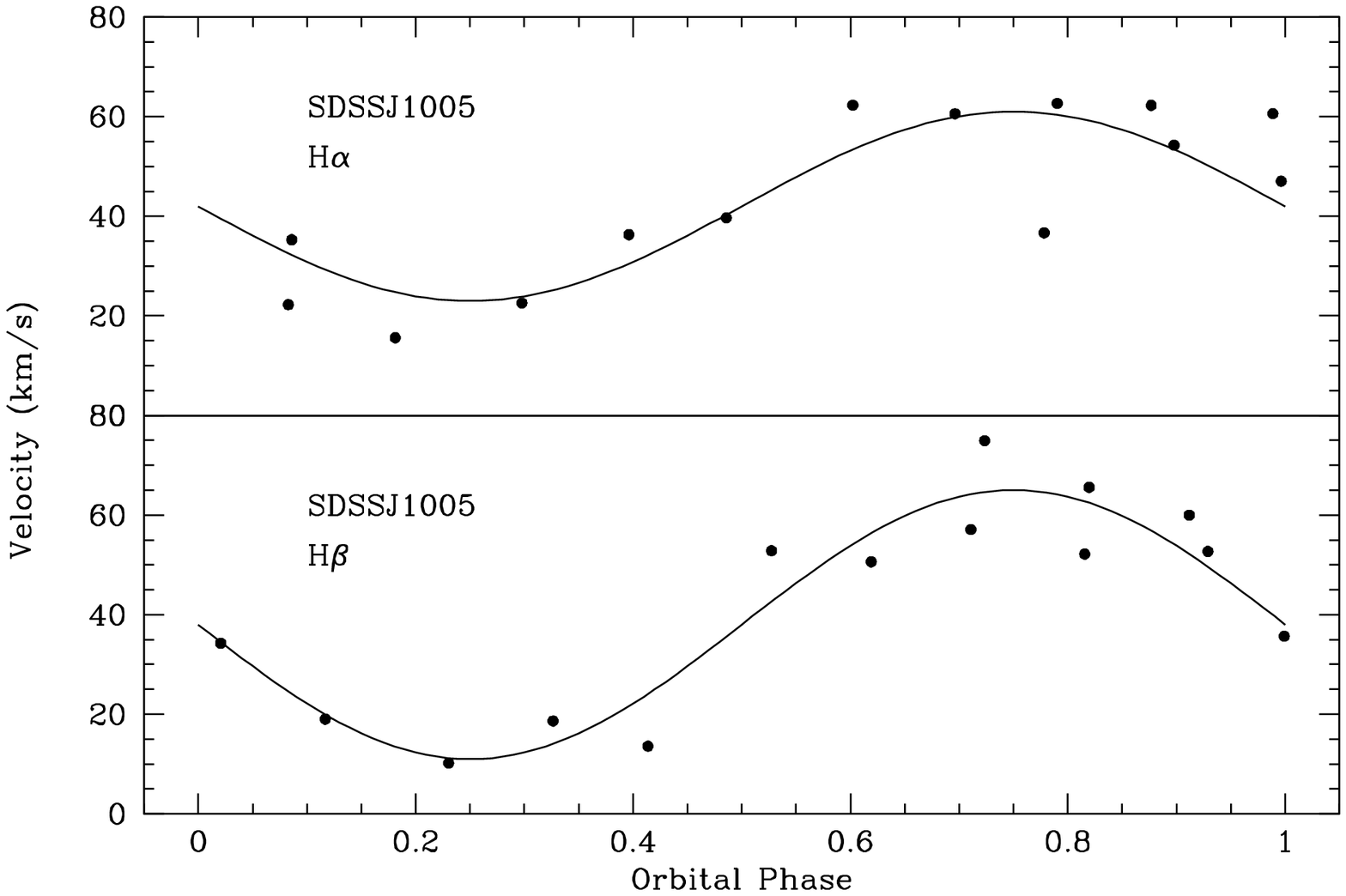}
\caption{H$\alpha$ and H$\beta$ velocity curves of SDSSJ1005 with the best fit sinusoids (Table 4) superposed. Sigmas of fits listed in Table 4.}
\end{figure}

\begin{figure} [p]
\figurenum {4}
\plotone{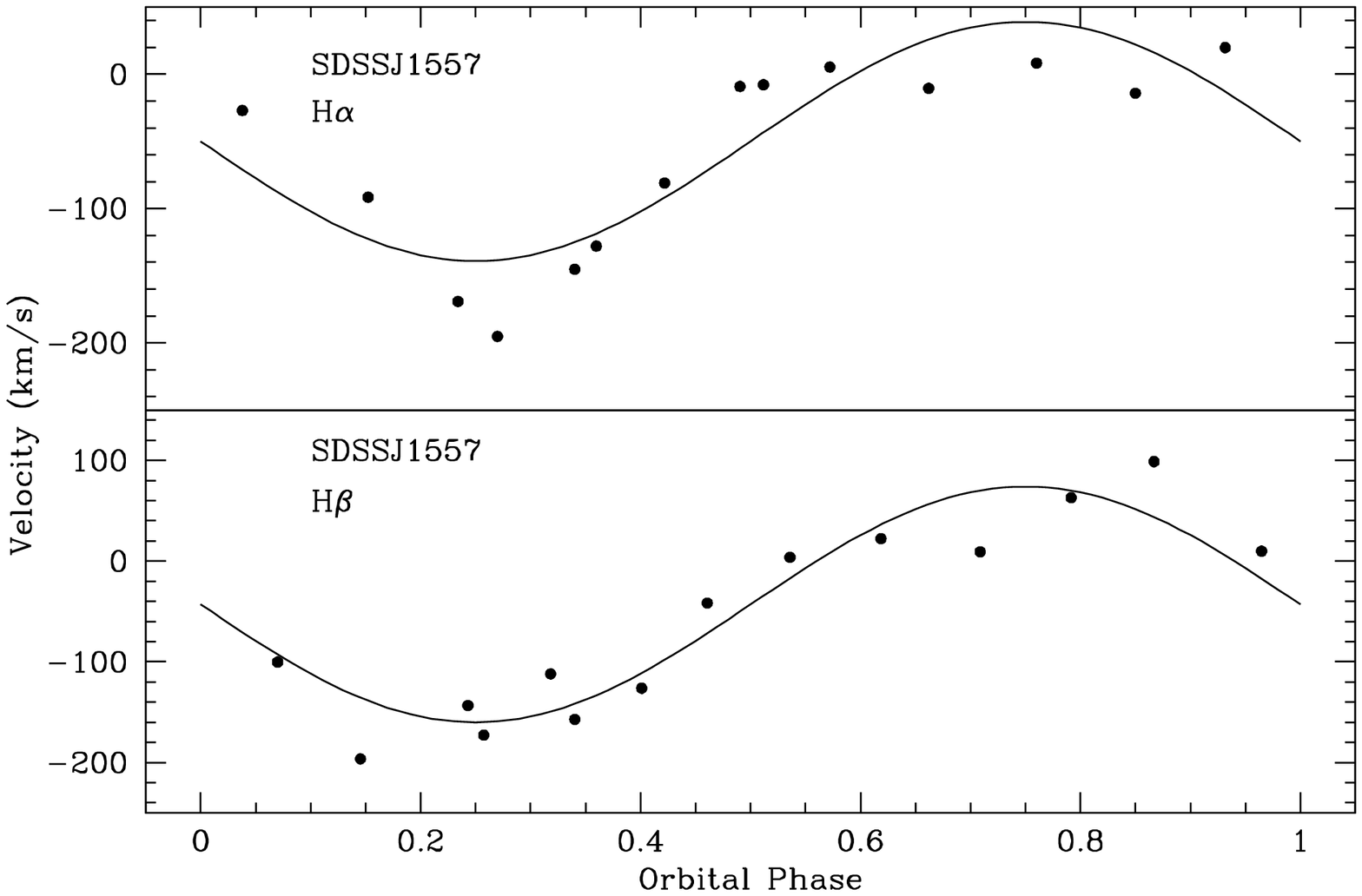}
\caption{H$\alpha$ and H$\beta$ velocity curves of SDSSJ1557 with the best fit sinusoids (Table 4) superposed.}
\end{figure}

\begin{figure} [p]
\figurenum {5}
\plotone{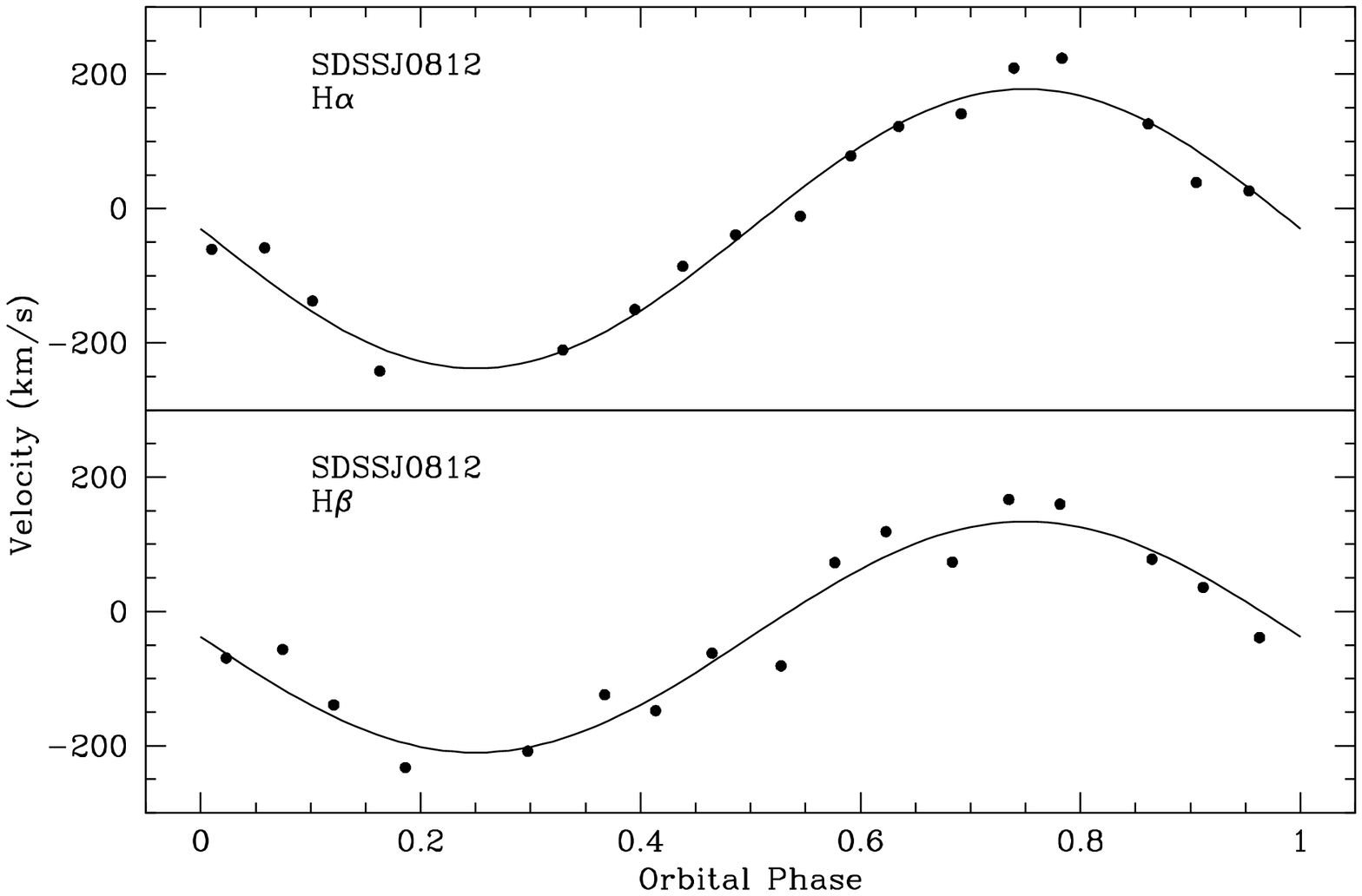}
\caption{H$\alpha$ and H$\beta$ velocity curves of SDSSJ0812 with the best fit sinusoids (Table 4) superposed.}
\end{figure}

\end{document}